\documentstyle[preprint,prl,aps]{revtex}

\begin{document}
\draft

\title{Exact Derivation of Luttinger Liquid Relation in a One-Dimensional
Two-Component Quantum System with Hyperbolic Interactions}

\author{Rudolf A.\ R\"{o}mer and Bill Sutherland}

\address{Physics Department, University of Utah, Salt Lake City, Utah 84112}

\date{February 15, 1994}
\maketitle

\begin{abstract}
We present an exact calculation of the Luttinger liquid relation for 
the one-dimensional,
two-component SC model in the interaction strength range $-1<s<0$ by
appropriately varying the limits of the integral Bethe Ansatz equations.
The result is confirmed by numerical and conformal methods. By a 
related study of the transport properties of the SC model, we can
give an exact formula for the susceptibility. 
Our method is applicable to a wide range of models such as, e.g., the
Heisenberg-Ising chain.
\end{abstract}


\narrowtext

%
%

The Luttinger liquid approach \cite{h8x} to one-dimensional quantum models in
conjunction with methods of conformal field theory \cite{bpz} is extremely
useful for understanding the low-energy physics of these models.
The essence of the method is the realization that close to the fermi
level, all excitations are essentially hydrodynamic density fluctuation
plus finite-size terms conveniently described by conformal finite size
formulas\cite{ca8x}.
Furthermore, the finite-size spectrum may be given in terms of a single
parameter due to the presence of the so-called Luttinger liquid relation.
In most models, this parameter is the renormalized interaction strength.

Although the set of models that fall into the Luttinger liquid universality
class (the $c=1$ universality class of the Gaussian model of conformal field 
theory) incorporates all ``classic'' short-ranged 1D models such as 
Heisenberg-Ising and Hubbard model\cite{ko93}, there is no rigorous proof that 
{\em all} possible 1D models should, too. 
Indeed, the calculation of the central charge $c$ in long-ranged models of
the $1/r^2$ type yields contradictory results\cite{ky91}.

In this Letter, we will give an exact derivation of a Luttinger liquid type
relation in a two-component quantum system with hyberbolic interaction defined
by the Hamiltonian
 \begin{equation}
H = - \frac{1}{2} \sum_{1\leq j \leq N} 
       \frac{\partial^{2}}{\partial x_{j}^{2}}
    + \sum_{1\leq j < k \leq N} v_{jk}(x_{j}-x_{k}),
\label{eqn-ham}
\end{equation}
with pair potential given as
\begin{equation}
v_{jk}(x) = s(s+1)
 \left[
  \frac{1+\sigma_{j}\sigma_{k}}{2 {\rm sinh}^{2}(x)}
  -
  \frac{1-\sigma_{j}\sigma_{k}}{2 {\rm cosh}^{2}(x)}
 \right], \quad s>-1.
\label{eqn-pot}
\end{equation}
The quantum number $\sigma=\pm 1$ distinguishes the two kinds of particles.
We may usefully think of it as either representing charge or spin. 
The model has been solved in Ref.\cite{sr93} by an asymptotic Bethe Ansatz
calculation and for large distances, its correlation functions have been 
calculated\cite{rs93}. We call it the SC model for the sinh-cosh interaction.
The SC model is closely related to the Hubbard and the
Heisenberg-Ising (H-I) model\cite{yy} and may be viewed as a H-I fluid,
with mobile spins\cite{rs94}. 

We restrict our investigation to $-1<s<0$, where the system has two
gapless excitations corresponding to a particle-hole and
a two spin-wave continuum with Fermi velocities $v_c$ and $v_s$,
respectively. 
Let $M$ be the number of spin-down
particles ($\sigma=-1$) and $N-M$ the number of spin-up particles for a total
of $N$ on a ring of length $L$. 
We then twist\cite{ss} the Bethe-Ansatz equations obtained in Ref.\cite{rs93} 
for $N$ pseudo-momenta ${\bf k}=(k_1, \ldots, k_N)$ and $M$ rapidities
$\mbox{\boldmath $\lambda$}=(\lambda_{1}, \ldots, \lambda_{M})$ with a flux $\Phi$ such that 
the total momentum $P({\bf k})=\sum k = 0$, i.e.,
\begin{mathletters}
\label{eqn-bakl}
\begin{equation}
L k 
 = 2\pi I(k) 
 - \frac{M}{N} \Phi 
 + \sum_{\lambda}^{M}\theta_{1}(k-\lambda)
 + \sum_{k'}^{N}\theta_{0}(k - k'), 
\label{eqn-bak}
\end{equation}
\begin{equation}
0 
 = 2\pi J(\lambda) 
 + \Phi 
 + \sum_{\lambda'}^{M}\theta_{2}(\lambda-\lambda')
 + \sum_{k}^{N}\theta_{1}(\lambda- k).
\label{eqn-bal}
\end{equation}
\end{mathletters}
The two-body phase shifts for particle-particle, particle-spin wave and 
spin wave-spin
wave scattering, i.e., $\theta_{0}(k)$, $\theta_{1}(k)$ and 
$\theta_{2}(k)$ respectively, have been given previously \cite{sr93}.
The particle quantum numbers $I_{j}$ and the spin-wave quantum numbers
$J_{a}$ are integers or half-odd integers depending on the parities 
of $N$, $M$ as well as on the particle statistics\cite{rs93}. 

In the thermodynamic limit, i.e., $L\rightarrow\infty$ at fixed particle
and spin wave densities, $d=N/L$ and $m=M/N$, respectively, ${\bf k}$ and
$\mbox{\boldmath $\lambda$}$ will be distributed densely around the origin.
The ground state is a filled
Fermi sea characterized by the distribution function $\rho(k)$ of particles
and $\sigma(\lambda)$ of down-spins.
so that we can replace the sums in Eq.(\ref{eqn-bakl}) with integrals. This
yields
\begin{mathletters}
\label{eqn-ibakl}
\begin{equation}
k 
 = \frac{2\pi I(k)}{L} 
 - \frac{M}{NL} \Phi 
 + \int_{-C}^{C} \theta_{1}(k-\mu) \sigma(\mu) d\mu
 + \int_{-B}^{B} \theta_{0}(k-h) \rho(h) dh, 
\label{eqn-ibak}
\end{equation}
\begin{equation}
0 
 = \frac{2\pi J(\lambda)}{L} 
 + \frac{\Phi}{L} 
 + \int_{-C}^{C} \theta_{2}(\lambda-\mu) \sigma(\mu) d\mu
 + \int_{-B}^{B} \theta_{1}(\lambda-h) \rho(h) dh,
\label{eqn-ibal}
\end{equation}
\end{mathletters}
and the values of $B$ and $C$ are fixed by the following equations:
\begin{eqnarray}
\int_{-B}^{B} \rho(k) dk &= &d,\\
\int_{-C}^{C} \sigma(\lambda) d\lambda &= &m d \equiv (d-y)/2.
\end{eqnarray}

We rewrite these equations using the operator notation of \cite{yy}:
$\langle k|\Theta_i|h \rangle\equiv\theta_i(k-h), 
 \langle k|\rho \rangle \equiv \rho(k)$, etc.. 
$B$ and $C$ are defined as
projection operators to yield the finite limits of the integration. 
In this notation, Eq.(\ref{eqn-ibakl}) is
\begin{mathletters}
\label{eqn-obakl}
\begin{equation}
k 
 = \frac{2\pi I(k)}{L} 
 - \frac{M}{NL} \Phi 
 + \Theta_{1}{C}\sigma
 + \Theta_{0}{B}\rho, 
\label{eqn-obak}
\end{equation}
\begin{equation}
0 
 = \frac{2\pi J(\lambda)}{L} 
 + \frac{\Phi}{L} 
 + \Theta_{2}C\sigma
 + \Theta_{1}B\rho.
\label{eqn-obal}
\end{equation}
\end{mathletters}

Let us take the derivative of Eq.(\ref{eqn-ibakl}) w.r.t.\ $k$. This then
yields a set of integral equations for $\rho(k)$ and $\sigma(\lambda)$
(Ref.\cite{rs93}, Eq.(10)). In terms of the operator notation we have
\begin{mathletters}
\label{eqn-odbakl}
\begin{equation}
1/2\pi 
 = \rho  
 + K_{1}{C}\sigma
 + K_{0}{B}\rho, 
\label{eqn-odbak}
\end{equation}
\begin{equation}
0 
 = \sigma 
 + K_{2}C\sigma
 + K_{1}B\rho,
\label{eqn-odbal}
\end{equation}
\end{mathletters}
where the new kernels are defined as 
$\langle k|K_i|h \rangle \equiv \frac{1}{2\pi} \theta_i'(k-h)$.

The neutral sector of the SC model is defined as $M=N/2$, i.e. $C=\openone$
and $y=0$.
We now wish to calculate the effect of a small disturbence $y\simeq 0$
($C\lesssim\openone$) and $\Phi\simeq 0$ on the system.
Eq.(\ref{eqn-odbal}) may be written as
\begin{eqnarray}
 \sigma 
 + K_{2}C\sigma
 &= &-K_{1}B\rho, \label{eqn-sigma} \\
 \sigma
 &= & -(1+J_2) K_1 B \rho + (1+J_2) K_2 (1-C) \sigma,
\end{eqnarray}
with $J_i$ the resolvent of $K_i$. Note that as pointed out in Ref.\cite{rs94},
the phase-shifts $\theta_1$ and $\theta_2$ are essentially the phase shift
of the H-I model after the identification of $|s|=\mu/\pi$. Therefore the
existence of the resolvents $J_1$ and $J_2$ in the regime $-1<s<0$ has been 
proved in Ref.\cite{yy}. 

Let $\sigma_0$ be a solution of Eq.(\ref{eqn-odbal}) with $C=\openone$, i.e.,
$\sigma_0 = -(1+J_2) K_1 B \rho$. Then we may also write
$
\sigma + J_2 (1-C) \sigma = \sigma_0
$.
We substitute this back into Eq.(\ref{eqn-ibak}) and have
\begin{equation}
k =
 \frac{2\pi I(k)}{L} 
 + \left[ \Theta_{0} - \Theta_{1}(1+J_2) K_1 \right] B \rho
 - \frac{M}{NL} \Phi
 - \Theta_1 (1+J_2) (1-C) \sigma. 
\end{equation}
Defining $\Theta\equiv\left[ \Theta_{0} - \Theta_{1}(1+J_2) K_1 \right]$
and $g\equiv -\frac{M}{NL} \Phi - \Theta_1 (1+J_2) (1-C) \sigma$ this yields
\begin{equation}
k =  
 \frac{2\pi I(k)}{L}
 + \Theta B \rho + g.
\label{eqn-pbak}
\end{equation}
We see that we have succeed in rewriting Eq.(\ref{eqn-bak}) as the
equation at $y=0$, $\Phi=0$ plus a small perturbation term $g$.

Let us briefly discuss how to handle the effect of this disturbance for
additive quantities such as the energy.
The net result of the presence of $g$ will be a shift in the pseudo-momenta
$k=k_0 + \delta k$, with $\delta k$ of order $1/L$.
Then, expanding Eq.(\ref{eqn-pbak}), eliminating terms of order $1$ with the
help of the unperturbed equation and defining $\gamma= \delta k \rho$, we have
$
\gamma + K B \gamma = g/2\pi
$.
Introducing the resolvent $(1+J)B(1+K)=(1+K)B(1+J)=\openone$, we can solve for
$\gamma$, i.e., 
$
\gamma = (1+J) B g/2\pi
$.
Let $E=\frac{1}{2} \sum^N k^2$ be the energy. Then the perturbed energy
to first order will be 
$E=\sum E(k_0 + \delta k)\simeq\sum E(k_0) +\sum E'(k_0)\delta k$ and we
calculate the difference due to the presence of the perturbation $g$ as
\begin{eqnarray}
\Delta E /L 
  &\equiv &(E(k) - E(k_0))/L \simeq B E' \delta k \rho
 \nonumber \\
  &=      &\int_{-B}^{B} E'(k) \gamma(k) dk = \gamma^{+} B E'
 \nonumber \\
  &=      &g^{+} B (1+J) B E' / 2\pi = g^{+} B (1+J) B k / 2\pi.
\label{eqn-de}
\end{eqnarray} 
The excitation spectrum of Bethe Ansatz models has been expressed by Yang
and Yang \cite{yye} in terms of the solution to an integral equation. For
the SC model the corresponding integral equation is
\begin{equation}
 \epsilon + K B \epsilon = k^2/2 - \mbox{(chemical potential)}.
\end{equation}
The chemical potential is chosen such that $\epsilon(\pm B)=0$ and thus we
may take the derivative of the above equation, perform a partial integration
and have
$
\epsilon' + K B \epsilon' = k$
or
$
\epsilon' = (1+J)B k
$.
Inserting this into Eq.(\ref{eqn-de}) and performing another partial integration,
we have the final result
\begin{equation}
\Delta E/L = - \epsilon^{+} B g'/2\pi = \epsilon^{+} B K_1 (1+J_2) (1-C)\sigma.
\label{eqn-del}
\end{equation}

In Eq.(\ref{eqn-bal}), $J(\lambda)$ was defined for $\lambda \in [-C,C]$ only.
However, the equations also work outside, i.e.,
$
0 
 = 2\pi J(\infty) + \Phi + (M-1) \theta_2(\infty) + N \theta_1(\infty)
 = 2\pi J(\infty) + \Phi + (M-1) \pi (1+2s)       - N \pi (1+s)
 \simeq 
   2\pi J(\infty) + \Phi + L [m d \pi (1+2s)       - d \pi (1+s)]
$.
Thus we have
\begin{eqnarray}
\int_{C_1}^{\infty} \sigma(\lambda) d\lambda
 &= & [J(\infty) - J(C_1)]/L \nonumber \\
 &\simeq & [\pi y (1+s) - (\Phi/L)]/2\pi,
\label{eqn-is}
\end{eqnarray}
since $J(C_1)=(M-1)/2\simeq L m d$. The same idea works for the lower limit.

Let us now return to Eq.(\ref{eqn-sigma}). We make the identifications
$\alpha = \pi\lambda$ and $\mu=\pi |s|$. Then the equation reads as
\begin{equation}
\sigma(\alpha)
+ \frac{1}{2\pi} \int_{-\pi C_2}^{\pi C_1} 
  \frac{\sin(2\mu)}{{\rm cosh}(\alpha-\beta)-\cos(2\mu)}
  \sigma(\beta) d\beta
= \frac{1}{2} \int_{-B_2}^{B_1} 
  \frac{\sin(\mu)}{{\rm cosh}(\alpha-\pi k)-\cos(\mu)}
  \rho(k) dk.
\end{equation}
If we compare this with Yang and Yang's \cite{yy} integral equation 
$R + {\bf K}R = \xi$,
we see that we can identify $\xi=-K_1(\alpha)/2$, ${\bf K}=K_2(\alpha)$ and write
\begin{equation}
\sigma(\lambda) = 
  \frac{1}{2} \int_{-B_2}^{B_1} 
   R(\pi\lambda - \pi k)
  \rho(k) dk.
\end{equation}
Furthermore, we have $\sigma_0 = R_0 B \rho/2 \simeq R_0 B \rho_0 /2$.
Again from Yang and Yang, we can then compute the asymptotic behavior
of $\sigma_0$, i.e.,
\begin{eqnarray}
\sigma_0 
 & \stackrel{\alpha\rightarrow\infty}{\longrightarrow}
 & \frac{\pi}{\mu} e^{-\frac{\pi\alpha}{2\mu}} 
    \frac{1}{2} \int_{-B_2}^{B_1} e^\frac{\pi^2 k}{2\mu} \rho_0(k) dk
 \nonumber \\
 & \simeq 
 & R_0(\alpha) a/2,
\end{eqnarray}
where $a$ is defined as the integral. Thus we write 
$\sigma(\lambda) \simeq a R(\pi\lambda)/2$.
Knowing the asymptotics of $R(\alpha)$ from Yang and Yang \cite{yy} and using
their notation, we can compute
\begin{eqnarray}
\int_{C_1}^{\infty} \sigma(\lambda) d\lambda
 &= & \frac{a}{2\pi} \int_{\pi C_1}^{\infty} R(\alpha) d\alpha \nonumber \\
 &= & \frac{a\zeta_1}{2\mu} \tilde{T}(0),
\end{eqnarray}
with $\zeta_1 \equiv e^{-\pi^2 C_1 /2\mu} \ll 1$.
The same calculation may be done for the lower limit and comparison with
Eq.(\ref{eqn-is}) yields
\begin{equation}
\zeta_{1,2} = \frac{\mu}{\pi a \tilde{T}(0)} 
 \left[ y(\pi-\mu) \mp (\Phi/L) \right].
\label{eqn-yphi}
\end{equation}
Using Eq.(\ref{eqn-del}), we may then express the corrections to the energy 
as
\begin{eqnarray}
\Delta E/L
 &= & -\case{1}{2}\epsilon^{+} B R_0 (1-C)\sigma \nonumber \\
 &= & -\case{1}{2} \int_B \epsilon (k) \left[
       \int_{1-C} R_0(\pi\lambda-\pi k) \sigma(\lambda) d\lambda \right]
       dk \nonumber \\
 &= & -\frac{\pi a}{4\mu} \int_B \epsilon (k) \left[
       \int_{1-C} e^{-(\pi^2\lambda-\pi^2 k)/2\mu} 
                  R(\pi\lambda) d\lambda \right] dk. 
\end{eqnarray}
We now again use the identification $\alpha=\pi\lambda$ and then shift
the integration boundaries from $\int_{C_1}^{\infty}$ ($\int^{C_2}_{-\infty}$)
to $\int_{0}^{\infty}$ ($\int^{0}_{-\infty}$). Furthermore, we treat the
contributions from the upper and lower limits as independent. Then we are
left with the Wiener-Hopf type problem of Yang and Yang \cite{yy} and we
may conveniently use their results, i.e.,
\begin{equation}
\Delta E/L
 =  -\frac{a}{4} \int_B \epsilon (k) e^{\pi^2 k/2\mu} dk
      \frac{1}{\pi} \left[ (\pi\zeta_1/\mu)^2 +(\pi\zeta_2/\mu)^2 \right]
      \tilde{T}(i\pi/2\mu).
\end{equation}
The $k$ integration may be rewritten by a partial integration
\begin{equation}
 - \int_B \epsilon (k) e^{\pi^2 k/2\mu} 
 = \frac{2\mu}{\pi^2} \int_B \epsilon' (k) e^{\pi^2 k/2\mu}
 \equiv  \frac{2\mu}{\pi^2} b,
\end{equation}
such that  
$
\Delta E/L = \frac{ab}{2\pi\mu} \tilde{T}(i\pi/2\mu) 
 \left[ \zeta_1^2 + \zeta_2^2 \right].
$
Using Eq.(\ref{eqn-yphi}), we then have
\begin{equation}
\Delta E/L =
 \frac{\tilde{T}(i\pi/2\mu)}{\tilde{T}^2(0)}
 \frac{b}{a}
 \frac{\mu}{\pi^3}
 \left[ (\Phi/L)^2 + (\pi-\mu)^2 y^2 \right].
\end{equation}
Again from Yang and Yang, we know 
$\frac{\tilde{T}(i\pi/2\mu)}{\tilde{T}^2(0)} = \frac{\pi^2}{8\mu(\pi-\mu)}$
and furthermore $b/a= 2\pi v_s$.
This gives the final result
\begin{equation}
\Delta E/L = 
 \frac{v_s (\pi-\mu)}{4} \left[ y^2 + \frac{(\Phi/L)^2}{(\pi-\mu)^2} \right].
\end{equation}
Defining the stiffness $D$ and the susceptibility $\chi$ as
$
\Delta E/L = \frac{1}{2} \left[ \chi^{-1} (y/2)^2 + D (\Phi/L)^2 \right],
$
we therefore have the desired relation
\begin{equation}
D \chi^{-1} = v_s^2.
\end{equation}

As we have shown in Ref.\cite{rs94}, this relation may be also derived from
conformal finite size formulas using the dressed charge matrix as
calculated by Wiener-Hopf techniques \cite{rs93,yy}. 
Furthermore, an analogous calculation yields the Luttinger 
relation for the H-I model and the method may be easily extended
to other models. However, we made good use of the seminal papers of
Yang and Yang \cite{yy,yye} in the present exposition and correspondingly 
useful results do not exist for most other models. We remark that
the essential idea for our calculation is already contained in these papers, 
and also in Haldane's ``Luttinger'' papers \cite{h8x}.

In Ref.\cite{rs94}, we have also shown how to relate stiffness and 
susceptibility 
by threading the system with a flux of $\Phi=2\pi(s+1)$. The Bethe Ansatz 
equations (\ref{eqn-bakl}) for $N$ particles and $M$ down-spins simplify 
considerably at this point and are in fact the equations for $N$ particles 
and $M-1$ down-spins at zero flux. This then gives 
$D=\chi^{-1}/4\pi^2(s+1)^2$. Thus we may express the susceptibility in terms
of the spin wave velocity as $\chi^{-1}= 2\pi v_s (s+1)$. In Fig.(\ref{fig-1})
we show a plot of $\chi^{-1}$ as function of the interaction strength. The
spin wave velocity has been calculated from the excitation spectrum of the
SC model as given in Ref.\cite{sr93}.
As $s\rightarrow 0^{-}$, the spin wave velocity approaches the velocity
of a non-interacting single-component model, i.e. $v_s\rightarrow \pi d/2$. 
Thus at half-filling, $\chi^{-1}(0^{-})$ approaches the non-zero value 
$\pi^2/2$ which is compatible with the result of Ref.\cite{ss}.

Finally, we wish to emphasize that we have derived the {\em exact} Luttinger
liquid relation in the SC model starting from the {\em asymptotic} Bethe
Ansatz equations. This further supports the use of the asymptotic Bethe Ansatz 
as an exact method in the thermodynamic limit.


R.A.R.\ gratefully acknowledges partial support by the Germanistic Society
of America.


\begin{figure}
  \caption{The inverse susceptibility $\chi^{-1}$ for the SC model from the
   Luttinger relation for $-1<s\leq 0$.
  \label{fig-1}}
\end{figure}

\end{document}